\documentclass[prb,twocolumn,showpacs,amssymb]{revtex4}
\usepackage{graphicx}

\begin{document}

\title{Antiferromagnetic domain walls in lightly doped layered cuprates \\}

\author{Baruch Horovitz}
\affiliation{Department of Physics and Ilze Katz center for
nanotechnology, Ben-Gurion University of the Negev, Beer-Sheva
84105, Israel}
\begin{abstract}
Recent ESR data shows rotation of the antiferromagnetic (AF) easy axis
in lightly doped layered cuprates upon lowering the temperature. We
account for the ESR data and show that it has significant
implications on spin and charge ordering according to the following
scenario: In the high temperature phase AF domain walls coincide with (110) twin
boundaries of an orthorhombic phase. A magnetic field leads to
annihilation of neighboring domain walls resulting in antiphase
boundaries. The latter are spin carriers, form ferromagnetic lines and
may become charged in the doped system.  However, hole ordering at low
temperatures favors the (100) orientation, inducing a $\pi/4$
rotation in the AF easy axis. The latter phase has twin boundaries and AF domain
walls in (100) planes.
\end{abstract}

\maketitle

The effect of holes on antiferromagnetism is a central issue in
high temperature superconductivity. In a remarkable recent experiment
the role of doping in
$Y_{1-x}Ca_xBa_2Cu_3O_6$ ($x\approx 0.008$) was studied
by J\'{a}nossy et al. \cite{yanossy} using an ESR technique. The
data shows that the antiferromagnetic (AF) polarization rotates
from [100] at high temperatures to [110] around $\approx 40K$. A
corresponding anomaly was also seen by $\mu$SR in $Y_{1-x}Ca_xBa_2Cu_3O_6$ as
well as in $La_{2-x}Sr_xCuO_4$ compounds \cite{niedermayer}. In the
undoped compound ($x=0$) the polarization remains in the [100]
directions down to low temperatures.

The ESR data \cite{yanossy} also shows that AF domains are present and that
their relative intensity is controlled by magnetic fields. In the high temperature
phase with field in the [100] direction, the [010] polarized
domains are preferred since the canting of AF spins has a higher
susceptibility. This defines a "depinning" field at which the
unpreferred domains are diminished. E.g. at high temperatures and above
a field of $\approx 1T$ 80\%
of the domains are polarized perpendicular to the field.
The ESR data shows, curiously, that
the depinning field of the [110] AF of $Y_{1-x}Ca_xBa_2Cu_3O_6$
(with field in the [1\={1}0] direction) is higher than that of the
[100] AF in $YBa_2Cu_3O_6$ (field in the [010] direction), both at low
temperatures.

In the present work we suggest that the transition in the AF
polarization correlates
with a lattice distortion and with condensation of the holes into
a charge density wave (CDW). Our reasoning involves a scenario for spin and charge
ordering with the following steps: (i) At
high temperatures and weak magnetic fields the [100] polarized AF
leads to an orthorhombic structure; the AF domain walls (DW's) coincide then
with twin boundaries at (110) planes. (ii) Strong magnetic fields
favor one of the AF domains so that DW's annihilate. We show that this
process results in antiphase boundaries (APB's) which consist of ferromagnetic
lines. (iii) At low temperatures hole ordering can occur by binding
states localized at the APB's. However, hole ordering  favors
the (100) orientation, i.e. a CDW with maxima on (100) planes and an
(incommensurate) wavevector in the [100] direction. This CDW favors an
AF  polarization in a diagonal [110] direction which is weakly orthorhombic with
(100) twin boundaries and DW's.

An AF polarized in the [100] axis favors an
orthorhombic phase since dipoles interact differently for $\rightarrow
\, \leftarrow$ and $\uparrow \, \downarrow$ pairs. This
magnetoelastic coupling is expected to be rather small in the
cuprates, leading to a change in lattice constant of order \cite{cimpoiasu}
$10^{-5}-10^{-6}$. In principle ESR analysis can
detect the resulting change in crystal field \cite{rettori},
however, the effect seems too weak in the case AF
$YBa_2Cu_3O_6$. Yet, this coupling can be sufficiently strong in samples
of size $\approx 1mm$ so that twin boundaries are necessary to relieve
macroscopic strains.

We proceed to present evidence for steps (i) , (iii) and then derive
step (ii). We claim that an orthorhombic structure is in fact supported by the
mere observation of AF domains. The presence of AF domains and the
associated domain walls \cite{farztdinov} are well known to result from either entropy
effects (when temperature is fairly close to the Neel temperature),
or from disorder or from a magnetoelastic coupling. The
ESR experiment \cite {yanossy} has used high purity samples and
temperature was well below the Neel temperature. Hence the presence
of DW's is a strong indication for the presence of the magnetoelasic
coupling, supporting step (i) of the scenario above.

Step (iii) is consistent with
data on the compounds \cite{fujita} $La_{1.875}Ba_{0.125-x}Sr_xCuO_4$.
These compounds maintain fixed carrier density
while allowing for both tetragonal and orthorhombic phases. In the
tetragonal phase an incommensurate CDW (as well as an
incommensurate spin density wave) with wavevector in the
[100] direction is present. For systems near the orthorhombic
boundary (less orthorhombic phase) the CDW wavevector deviates
slightly from the [100] axis and finally the CDW disappears in the
orthorhombic phase.

We propose then that the orthorhombic structure induced by the
[100] polarized AF structure is unstable when holes are
added and form a [100] CDW. The AF polarization then rotates to
the [110] direction which, as shown below, is weakly orthorhombic and has (100) twin
boundaries. We proceed now to
analyze an effective free energy and study various domain walls.
In particular step (ii) is shown, i.e. the collapse of two
neighboring magnetic DW's (e.g. by applying a magnetic field) leads
to antiphase boundaries which can be detected by a variety of ¤experiments.

\begin{figure}
\begin{center}
\includegraphics[scale=0.5]{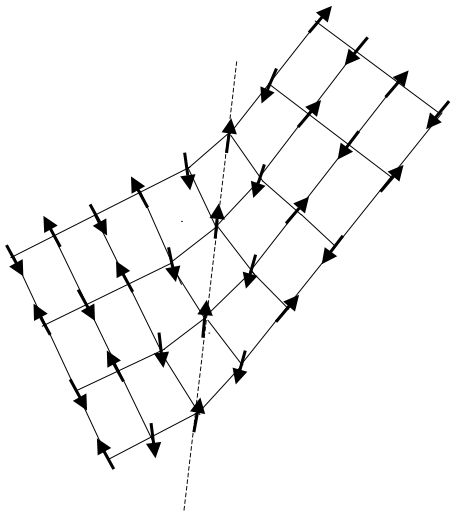}
\end{center}
\caption{Twin boundary with spin polarizations (arrows) exhibiting
an AF domain wall. The dashed line is in a (110) plane and is the $s=0$ line defining a
TB position. The displacement $u(s)$ is parallel to this line and
yields the strain $\epsilon=\partial u/(\sqrt{2}\partial s)$
interpolating between the two orthorhombic variants.\vspace{3mm}}
\end{figure}

 Consider first the elasticity part, i.e. the
tetragonal-orthorhombic (T-O) transition and formation of twin
boundaries (TB's). The orthorhombic distortion, for weak strain,
corresponds to a tetragonal lattice displaced in the [110]
direction with a displacement $u$ that is linear in the coordinate
$s$ in the [1\={1}0] axis, i.e. $u\sim \pm s$ (see Fig. 1). The $\pm$
signs correspond to the two variants of the orthorhombic phase.
The free energy, in terms of the strain $\epsilon=\partial
u/(\sqrt{2}\partial s)$, has then the form \cite{barsch}
\begin{equation}\label{F0}
{\cal F}_0={\cal F}_1(\epsilon) + \frac{1}{2}c
\left(\frac{\partial \epsilon}{\partial s}\right)^2
\end{equation}
where ${\cal F}_1(\epsilon)$ has a double minima corresponding to
the two variants. A static solution interpolating smoothly between
the two variants is then possible \cite{barsch}, and is described
by a function $\epsilon_{TB} (s)$. This function has necessarily
an $s$ value where $\epsilon_{TB} (s)=0$ which defines a twin
boundary location, hence the TB is on a (110) plane. In
general, a TB orientation is uniquely determined by the type of
structural transformation, i.e. smooth interpolation between
variants across other planes leads to diverging elastic energies
\cite{khach}. The formulation of Eq. (\ref{F0}) is a simple
demonstration that for the T-O transition TB's are on (110)
planes. A macroscopic strain which maintains the overall sample shape, e.g.
by presence of a parent (tetragonal) phase\cite{horovitz}, imposes a TB
array and determines its periodicity.

We proceed now to study the T-O transformation induced by a [100]
polarized AF as well as the possible AF domain walls coexisting
with TB's. An AF domain wall is defined by a localized rotation of
the polarization angle $\theta$ (e.g. relative to the [100] axis)
by $\pm\pi/2$, i.e. neighboring domains are polarized along [100]
and [010], respectively. The coupled AF and strain structure is
illustrated in Fig. 1. Note that antiparallel spins favor
shorter bonds, therefore a TB induces a $\pi/2$ rotation in the AF
polarization, as shown in the figure.

We claim that neighboring DW's have the same rotation angle
$\pi/2$ (or $-\pi/2$), rather than opposite $\pi/2$ and $-\pi/2$.
To show that the latter has no solution note first that the
macroscopic strain which imposes the TB array does not affect the
local differential equation for $\theta(s)$, instead it determines
the average TB spacing\cite{horovitz} being an integration
constant. Hence the existence of a static solution $\theta(s)$ can
be deduced from dynamical stability considerations of the AF
structure by itself. In particular DW's with opposite $\pi/2$ and
$-\pi/2$ rotations are unstable since by approaching each other
they can annihilate and form
the lower energy ground state; in contrast, two $\pi/2$ DW's when
approaching each other form a higher energy antiphase boundary
(see below), hence they are stable and a static solution is
possible.

To illustrate this idea more explicitly we consider a slowly
varying magnetization with a magnetoelastic coupling energy of the
form $-a\epsilon \cos (2\theta)$; in particular a [110]
polarization ($\theta=\pi/4$) does not couple to this strain (the
$\pi/4$ state has a weaker coupling to a different strain, as
considered below). $F_1$ in Eq. (\ref{F}) is replaced now by a
quadratic term $+\frac{1}{2}b\epsilon^2$ since the T-O transition
is driven by the magnetoelastic coupling, hence the full free
energy is
\begin{equation}\label{F}
F= -a\epsilon \cos (2\theta) +\frac{1}{2}b\epsilon^2 +
\frac{1}{2}c\left(\frac{\partial \epsilon}{\partial s}\right)^2 +
\frac{1}{2}d\left(\frac{\partial \theta}{\partial s}\right)^2
\end{equation}
where the constants $b,c,d$ are positive and the last term
represents the spin stiffness. (A crystal field
$\sim \cos (4\theta)$ from spin orbit coupling is neglected; in fact
it must be small to ensure that $\theta=0$ is a ground state.)
The two orthorhombic variants correspond to $\theta=0, \pi/2$
with the strain $\epsilon=\pm a/b$.

Domain walls are solutions of the minimum condition for Eq.
(\ref{F}),
\begin{eqnarray}
-a\cos (2\theta)+b\epsilon-c\frac{\partial^2 \epsilon}{\partial^2
s}&=&0  \nonumber  \\
-2a\epsilon \sin (2\theta)+d\frac{\partial^2 \theta}{\partial^2
s}&=&0 \label{min2}\,.
\end{eqnarray}
A TB which interpolates between
the variants $\epsilon=\pm a/b$ necessarily leads to rotation of
the AF polarization by $\pi/2$, as in Fig. 1. We show next that a
periodic TB array for which $\theta(s)$ is a monotonic
function is a valid solution. Consider a segment of such a TB
array where $\theta=0$ at $s=-s_0$, becomes $\theta=\pi/4$ at
$s=0$ and finally rotates to $\theta=\pi/2$ at $s=s_0$; in the
same interval the strain is antisymmetric, interpolating from
$\epsilon$ near $a/b>0$ (assuming $a>0$) to $\epsilon$ near
$-a/b<0$. Integration of Eq. (\ref{min2}) yields
\begin{equation}\label{min3}
\frac{\partial \theta (s)}{\partial s}-\frac{\partial \theta
(s)}{\partial s}|_{-s_0}=\frac{2a}{d}\int_{-s_0}^s\epsilon(s)\sin
[2\theta(s)]ds \,.
\end{equation}
In the interval $-s_0<s<s_0$ $\sin (2\theta(s))>0$ is symmetric
while $\epsilon$ is antisymmetric starting from positive values,
hence the right hand side of Eq. (\ref{min3}) is positive and
approaches zero at $s_0$. At the next TB, $\epsilon$ is
antisymmetric starting from {\em negative} values while $\sin
(2\theta(s))$ is symmetric and negative, hence again the right
hand side of Eq. (\ref{min3}) is positive. We conclude that
$\partial \theta/\partial s>0$ is a consistent solution. Furthermore, in
the limit $c\ll d$ the gradients are dominated by the spin
stiffness while the strain follows the local AF angle, $\epsilon
\approx (a/b)\cos (2\theta)$ and Eq. (\ref{min2}) becomes
$-(a^2/b)\sin (4\theta)+d
\partial^2 \theta/\partial^2 s=0$. The latter is the sine-Gordon
equation with well known periodic domain structure satisfying
$\partial \theta/\partial s>0$.

\begin{figure}
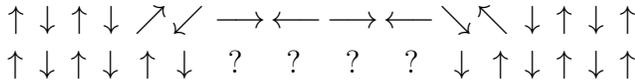

\begin{large}
\begin{eqnarray}
\uparrow \,\,\, \downarrow \,\,\, \uparrow \,\,\, \downarrow
\,\,\,
  \nearrow \,\swarrow \,\,
\longrightarrow \,\longleftarrow \,\longrightarrow
\,\longleftarrow \,\searrow \,\nwarrow \,\,\,\downarrow
\,\,\,\uparrow \,\,\,\downarrow \,\,\,\uparrow &&
 \nonumber \\
  \uparrow \,\,\, \downarrow \,\,\, \uparrow \,\,\, \downarrow
\,\,\,\, \uparrow \,\,\,\, \downarrow \;\;\;\; ? \;\;\;\;\; ?
\;\;\;\;\; ? \;\;\;\;\; ? \;\;\;\downarrow \;\;\,\uparrow
\,\,\,\downarrow \,\,\,\uparrow \,\,\,\downarrow \,\,\,\uparrow
&&\nonumber
\end{eqnarray}
\end{large}
\caption{The 1st line shows three AF variants along [100],
separated by two DW's (with the diagonal polarizations). The 2nd
line shows the result of annihilation of the two DW's -- the spins
marked by ? continue either left or right domains and an antiphase
boundary results somewhere in the ? marked range.\vspace{3mm}}
\end{figure}

A $\pi/2$ DW is therefore followed by the {\em same} sign $+\pi/2$
DW. This result is, in fact, a robust consequence of the energy
argument above requiring DW's stability, hence it is valid even
with additional terms in Eq. (\ref{F}). A remarkable consequence
of this analysis is that when two neighboring TB's annihilate the
result is not a uniform ground state, but rather a change of
$\theta(s)$ by $\pi$ which is an antiphase boundary (APB), i.e. all spins
on one side of the boundary are reversed relative to the ground state
orientations. The
annihilation process of DW's is shown in Fig. 2. The 1st line
shows the spin arrangement of three variants separated by two DW's
with the polarization angle increasing monotonically. In the 2nd
line the DW's are eliminated, i.e. the spins are rotated so as to
be aligned with either the left or the right domain. The diagonal
spins join their nearest domain, but the horizontal ones can
choose either domain (indicated by a ? mark), with any choice
resulting in an APB somewhere in the ? marked region. Note that
the APB carries an additional spin 1/2 relative to the AF ground
state.

APB's involve a major rearrangement of the electronic coordinates
and can result in a localized state, as shown by a mean field
study \cite{zaanen}, analogous to studies on solitons in polyacetylene.
An APB involves therefore atomic scales and its width is of that
order. In contrast, an AF DW as well as a TB, have a much larger
width and therefore lower formation energies. The high energy cost of
an APB, relative to that of two DW's, provides the stability of the
DW array with same sign DW's.

DW's can be eliminated by a strong magnetic field as was
demonstrated by the experiment of  J\'{a}nossy et al.\cite{yanossy}, leading to
a single variant. The process can generate APB's assuming that
they are sufficiently apart and do not annihilate each other.
APB's carry spin 1/2 per $Cu$ along the boundary, i.e. a
ferromagnetic line along [110] as shown in Fig. 3. These lines are
AF in the [001] direction, yet, the interlayer coupling is weak so
that it is relatively easy to flip magnetization by a field {\em
parallel} to the AF polarization. Hence, when the magnetic field
overcomes the interlayer coupling a nonlinear longitudinal
susceptibility $\chi_{\parallel}$ appears due to aligned
ferromagnetic lines. Note that by rotating the APB of Fig. 3 into a (100)
plane its lines (within the layers) are AF rather then ferromagnetic.

\begin{figure}
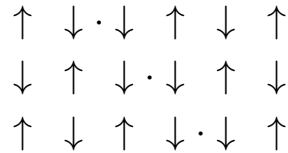

\begin{Large}
\begin{eqnarray}
\uparrow \;\;\; \downarrow \cdot \downarrow \;\;\; \uparrow
\;\;\; \downarrow  \;\;\; \uparrow && \nonumber \\
\downarrow \;\;\; \uparrow \;\;\; \downarrow \cdot  \downarrow
\;\;\; \uparrow  \;\;\; \downarrow && \nonumber \\
\uparrow \;\;\; \downarrow \;\;\; \uparrow \;\;\; \downarrow \cdot
\downarrow  \;\;\; \uparrow && \nonumber
\end{eqnarray}
\end{Large}
\caption{An APB along a (11) line, centered on the dots, showing a
ferromagnetic line. Note that the ferromagnetic polarization is
parallel to that of the AF.\vspace{3mm}}
\end{figure}

\begin{table*}
\caption{ The table lists topological structures in the AF
ordering in lightly doped Cuprates, as well as pertinent
experimental signatures at various magnetic fields. High
temperature corresponds to the [100] polarized AF phase and low
temperature corresponds to the [110] polarized AF with a [100] CDW
phase.}
\begin{ruledtabular}
\begin{tabular}{|c|c|c|}
& high temperature   & low temperature \\
&   &  \\ \hline
strong [100] field & (110) neutral APB & (100) DW and (100) charged APB \\
&  ferromagnetic lines, nonlinear $\chi_{\parallel}$  & \\ \hline
strong [110] field & (110) DW  & (100) charged APB \\
& & higher $T_{CDW}$ \\ \hline
no field &  (110) DW   &   (100) DW and (100) charged APB  \\
&   &   stronger depinning than for the undoped case  \\
\end{tabular}
\end{ruledtabular}
\end{table*}

Consider next an alternative AF phase with polarization along [110],
stabilized by terms beyond those in Eq. (\ref{F}), e.g. CDW ordering
(see below). The coupling of nearest neighbors along [100] or [010]
is now identical and orthorhombicity is not induced, i.e. the first
term in (\ref{F}) vanished at $\theta=\pi/4$. However, second nearest
neighbors form either $\uparrow, \uparrow$ or $\rightarrow,
\rightarrow$  pairs which are nonequivalent. Hence a tetragonal unit
cell which is formed by the vectors [110], [1\={1}0], [001] of the
original lattice tends to contract or expand along [110] and
[1\={1}0], respectively. This unit cell is rotated by $\pi/4$ relative
to the previous one, hence the axis $s'$ along which an orthorhombic
distortion alternates is now in the [100] direction, therefore twin
boundaries that interpolate between $\theta=\pi/4$ and
$\theta=3\pi/4$ are in (100) planes of the original tetragonal lattice.
The detailed description of
these TB's is more involved than the previous ones since near
$\theta\approx\pi/2$ the previous strain will couple. However,
following the stability argument above, we expect again a
$\pi/2$ DW to be followed by the {\em same} sign $+\pi/2$ DW, hence
 a strong magnetic field in a [110]
direction would coalesce these TB's and lead to (100) APB's.

The final ingredient in our model is that the added holes form a
CDW with wavevector along [100] as clearly seen in the
$La_{1.875}Ba_{0.125-x}Sr_xCuO_4$ compounds \cite{fujita}. Indeed
the charged APB in the mean field calculation \cite{zaanen} is more strongly
bound when the APB is along a (100) plane. These (100) APB's form
a periodic array of stripes equivalent to a CDW with wavevector in the [010]
direction. In a [100] polarized AF these (100) APB's would cross (110) DW's or
(110) APB's; the [100] CDW therefore favors a rotated [110] polarized AF so that
crossings with DW's are avoided. Furthermore, in a strong [110] field
 the [110] polarized AF has the
appropriate ingredients for forming stripes, namely (100) APB's,
facilitating the CDW formation. We expect therefore that the
transition temperature $T_{CDW}$ into a CDW be higher with a field in
this [110] direction. The predictions for DW's and APB's are summarized
in table I, as well as pertinent experimental signatures.

We show now that our model accounts also for the observation that
depinning field in the [110] polarized AF $Y_{1-x}Ca_xBa_2Cu_3O_6$ is larger
than that of the [100] polarized AF in $YBa_2Cu_3O_6$, both at low
temperatures \cite{yanossy}. Depinning involves annihilation of DW in pairs
leading to a single domain. In $Y_{1-x}Ca_xBa_2Cu_3O_6$ some of
the DW need to cross a charged APB, a
process which has a barrier. In $YBa_2Cu_3O_6$ the process
involves TB annihilation which is continuously achieved as field
is increased with no barrier. Hence a larger depinning field in
$Y_{1-x}Ca_xBa_2Cu_3O_6$.

Finally, we propose a variety of experiments for observing APB's and
probing our scenario:
(i) APB's are expected to have intragap states \cite{zaanen},
hence optical absorption should show new lines at high magnetic
fields. (ii) The ferromagnetic nature of (110) APB lines (Fig. 3)
results in a nonlinear longitudinal susceptibility $\chi_{\parallel}$. (iii) By adding holes
the APB intragap state can be charged, leading to a spinless
charge carrier. These charges affect the optical absorption in (i)
as well as the susceptibility in (ii). (iv) The CDW onset is
facilitated by a strong magnetic field in the [110] direction, hence
$T_{CDW}$ is higher in this case.

In conclusion, our scenario accounts for the unusual ESR data \cite{yanossy}
and predicts a variety of topological structures (table I). We propose that
experiments based on these observations can supplement the
ESR data as well as clarify a central issue in high temperature
superconductivity -- the nature of doping in AF layered cuprates.

\vspace{3mm}
Acknowledgements: I thank Prof. Andr\'{a}s J\'{a}nossy for
stimulating discussions and for his hospitality, and the Institute
of physics at the Budapest University of Technology and Economics
for hospitality during this research. I also thank D. Golosov for
pointing out Ref. \onlinecite{farztdinov} and A. Aharony for useful
comments. This research was supported
by THE ISRAEL SCIENCE FOUNDATION founded by the Israel Academy of
Sciences and Humanities.

\end{document}